\documentclass[a4paper,11pt,table]{article}
\usepackage{pos}

\title{Core-Collapse Supernova Search Strategy: Gravitational Waves and Low-Energy Neutrinos}

\author*[a]{Odysse Halim}
\author[b]{Claudio Casentini}
\author[c,d]{Marco Drago}
\author[e,f]{Viviana Fafone}
\author[g]{Kate Scholberg}
\author[h,i]{Carlo Vigorito}
\author[j,k]{Giulia Pagliaroli}

\affiliation[a]{Istituto Nazionale di Fisica Nucleare (INFN) sez. di Trieste, Italy,}

\affiliation[b]{Istituto Nazionale di Astrofisica - Istituto di Astrofisica e Planetologia Spaziali (INAF -IAPS), Rome, Italy,}

\affiliation[c]{Universita` di Roma La Sapienza, I-00185 Roma, Italy,}
\affiliation[d]{INFN, Sezione di Roma, I-00185 Roma, Italy}
\affiliation[e]{University of Rome Tor Vergata, Rome, Italy,}
\affiliation[f]{INFN sez. di Roma Tor Vergata, Rome, Italy,}
\affiliation[g]{Department of Physics, Duke University, Durham, NC, USA,}
\affiliation[h]{University of Turin, Italy,}
\affiliation[i]{INFN sez. di Torino, Italy,}
\affiliation[j]{Gran Sasso Science Institute (GSSI), L’Aquila, Italy,}
\affiliation[k]{INFN sez. di LNGS, Assergi, Italy}

\emailAdd{odysse.halim@ts.infn.it}

\abstract{Core-collapse supernovae are expected to produce multimessenger signals. Low-energy neutrinos and gravitational waves are important to study the explosion mechanism of these events. The simulations and detections of gravitational waves from these events are still challenging due to broad range of expected progenitors as well as their stochasticity. In this manuscript, we give a timely introduction to a proposed method to combine gravitational waves and neutrinos from core-collapse supernovae; more detailed discussion will be provided in a forthcoming paper \cite{Halim2021} (\href{https://arxiv.org/abs/2107.02050}{arXiv:2107.02050}). We discuss how to exploit the time profile of the neutrino signals in order to improve the efficiency search. Moreover, we describe the combination of neutrino data from several detectors. The goal is to produce a strategy for combining the neutrinos and gravitational waves as a multimessenger search.}
\FullConference{%

European Physical Society conference on High Energy Physics (EPS-HEP) 2021\\
July 26-30, 2021\\
\href{https://www.eps-hep2021.eu/}{Online conference}
}

\begin{document}
\maketitle

\section{Introduction}

Core-collapse supernovae (CCSNe) are expected to produce multimessenger signals \cite{pagliaroli_PRL,leonor}. Low-energy neutrinos (LENs) with average energy $\sim 10$ MeV are expected to be copiously emitted by these events, from the majority of the total energy budget ($\sim 10^{53}$ erg), which was the case of the last CCSN in the nearby galaxy, SN1987A detection \cite{hirata1987,Bionta1987,alexeyev}. Gravitational waves (GWs) and multi-wavelength electromagnetic emissions will also be produced by CCSNe. 

\begin{figure}[!ht]
\centering
\includegraphics[width=.6\linewidth]{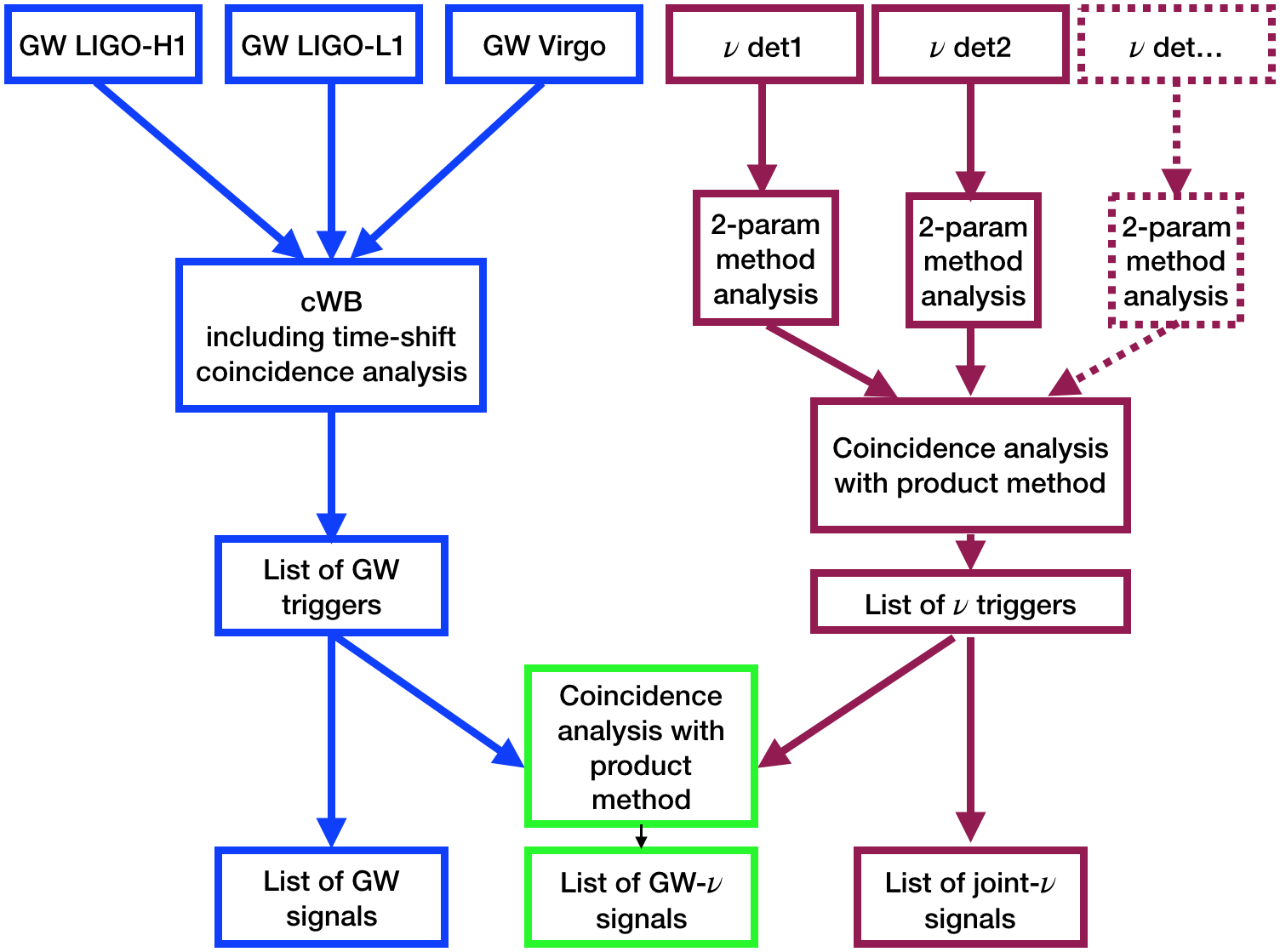}
\caption{The schematic view of the multimessenger GW-LEN strategy.}
\label{fig:GWnu_scheme}
\end{figure}

Here, we show a timely introduction of our method to combine GWs and LENs from CCSNe; more details will be published soon \cite{Halim2021} (already accepted by JCAP). We construct a strategy in order to combine GWs and LENs for a multimessenger search. We use \texttt{coherent WaveBurst} (cWB) pipeline \cite{Klimenko_2004,Klimenko2008,dragotesi,Necula_2012} for unmodelled GW data analysis from simulations. Moreover, we also simulate LEN signals arriving in several neutrino detectors and analyse them. At this stage, for neutrino analysis, we exploit the temporal behaviour of CCSNe. In the next step, we do a temporal-coincidence analysis between two messengers following the chart in Fig.~\ref{fig:GWnu_scheme}. Each messenger data is analysed separately and then combined together for possible GW-LEN signals. This could be interesting for online networks such as SNEWS or offline search for sub-threshold signals. This paper will start on a discussion of the emission models from each messenger. Then, the data and analysis by our strategy will be presented. In the end, we show the result.

\section{Messengers from Core-collapse Supernovae}\label{sec:emission}
A CCSN is expected to produce a $O(10)$-ms GW burst and an impulsive $O(10)$-sec emission of  $O(10)$-MeV LENs \cite{jankarev}. Besides, these signals could also be emitted by ``failed'' SNe \cite{OConnor:2010moj}. GW and LEN signals highly depend on the initial conditions of the CCSN, as progenitor mass, rotation, etc. The ideal case is to perform a coherent GW-LEN analysis by considering simulated GW and LEN signals resulting from the same numerical simulation. Unfortunately, no coherent GW-LEN numerical simulations providing successful CCSN explosion exist up to now. Several simulations provide both signals for the first half of a second, till the explosion, while the LEN emission duration is expected to last $O(10)$-sec. Therefore, we use the presently available GW signals and neutrino signals from different simulations with similar progenitor masses, and treat them as if they are from the coherent simulation.



The detail of our GW waveforms are shown in c.f. Tab.~1 in \cite{Halim2021}. We consider the GW signals from the 3D neutrino-radiation hydrodynamics simulations of Radice {\it et al.} \cite{radice} (named as ``Rad'') for various zero age main sequence masses in order to take into account the successful explosions from the low-mass and the failed explosions from the high-mass. We also take into account models with rapid rotation and high magnetic field. In this case, we take GW waveforms from two simulations: the Dimmelmeier \cite{dimmelmeier2008} (named as ``Dim'') and the Scheidegger \cite{Scheidegger:2010en} (named as ``Sch''). These simulations produce much stronger GWs, however, the stellar progenitors must have strong rotation and magnetic field, expected to be less probable than the neutrino-radiation mechanism \cite{em_gw_ccsne,jankarev,woosley}. Nevertheless, we have not yet detected any CCSN GWs from any possible models so we would not rule out any models.









Concerning the LEN emissions, we consider two models. The first one is the signals from the numerical simulations of H{\"u}depohl without the collective oscillations \cite{hudepohl}. We adopt the time-dependent neutrino luminosities and average energies obtained for a progenitor of $11.2 M_\odot$. The simulation provides the first 7.5 seconds of the neutrino emission and we make an analytical extension in order to reach 10 seconds of the signal. The average neutrino energies from before collapse up to the simulated $0.5$ s after bounce are (c.f. Table 3.4 of Ref. \cite{hudepohl}): $\langle E_{\nu_e}\rangle=13$ MeV, $\langle E_{\bar{\nu}_e}\rangle=15$ MeV and $\langle E_{\nu_x}\rangle=14.6$ MeV. The second model is a parametric model for neutrino emission from Pagliaroli {\it et al} \cite{pagliaroli2009}. This model provides the best-fit emission from SN1987A data with the average energies of $\langle E_{\nu_e}\rangle=9$ MeV, $\langle E_{\bar{\nu}_e}\rangle=12$ MeV 
and $\langle E_{\nu_x}\rangle=16$ MeV.




\section{Data and Analysis Strategy\label{sec:science}}

Here, we will discuss in brief the data and analysis for GWs, for LENs, as well as a possible strategy to do a combined multimessenger search in the context of our work. More details can be seen in \cite{Halim2021}. We consider a conservative threshold on global false alarm rate (FAR) of 1/1000 years. 

 

The GW analysis uses the cWB algorithm, a pipeline that has been used widely inside the LIGO and Virgo collaborations. For this work, we simulate Gaussian noise with a spectral sensitivity based on the expected Advanced LIGO \cite{TheLIGOScientific:2014jea,Abbott:2020qfu} and Advanced Virgo detectors \cite{TheVirgo:2014hva}. We use about 16 days of data and around 20 years of background livetime. Waveforms from the emission models in Sec.~\ref{sec:emission} have been generated for distances: 5, 15, 20, 50, 60, 700 kpc. We inject 1 GW signal in about 100 seconds, wide enough to allow two consecutive waveforms. To ensure sufficient statistics, for each distance and model we inject $\sim 2500$ different realizations over all the sky direction. We apply a threshold of $864$ per day for GW candidates \cite{Halim2021} to pass to the multimessenger analysis, as per requirement of the combined FAR of 1/1000 years. The efficiency curves for each distance can be seen in c.f. Fig.~2 of \cite{Halim2021}





In the standard LEN analysis for CCSNe, a time series data set is binned in a sliding time window of $w=20$ seconds. The group of events in each bin is defined as a \textit{cluster} and the number in the cluster is called multiplicity $m$. The multiplicity distribution of background-only events is expected to follow a Poisson distribution. For each $i$-th cluster, the significance can be estimated. The significance is correlated with its \textit{imitation frequency} ($f^\mathrm{im}$, more detail in \cite{Halim2021}) defined as:
\begin{equation}
f^\mathrm{im}_i (m_i)=N\times \sum_{k=m_i}^\infty P(k),
\label{eq:fim_prob}
\end{equation}
where $P(k)$ is the Poisson term, representing the probability that a cluster of multiplicity $k$ is produced by the background with a given frequency and time window.         



In our case, we exploit the temporal behaviour of CCSNe by defining a new parameter $\xi\equiv m/\Delta t$, where $\Delta t$ is the duration of a cluster (see \cite{Halim2021}). Thus, it is formulated a new modified \textbf{2-parameter} ($m$ and $\xi$) imitation frequency for each cluster, called $F^\mathrm{im}$:
\begin{equation}
F^\mathrm{im}_i (m_i,\xi_i)=  N \times  \sum_{k=m_i}^\infty  P(k,\xi_i),
\label{eq:newfim_0}
\end{equation}
where the term $P(k,\xi_i)$ provides the joint probability of a cluster, given multiplicity $k$ \textbf{and} $\xi_i$, produced by the background. The final form can be seen in Eq.~5 of \cite{Halim2021}, which is the main equation in this work,

\begin{equation}
F^\mathrm{im}_i(m_i,\xi_i)= N \times \sum_{k=m_i}^\infty P(k) \int_{\xi=\xi_i}^\infty \mathrm{PDF}(\xi\geq\xi_i|k) d\xi.
\label{eq:newfim_1}
\end{equation}

We simulate $\sim10$ years of background data for each neutrino detector. Moreover, the CCSN simulated signals are injected into the background data for each detector model and for each source distance coherently in time. These clusters of neutrino events are extracted via the Monte Carlo method. We inject signals at a rate of 1 per day.

The ultimate goal of this paper is to construct a multimessenger analysis, combining both LEN and GW data sets (green boxes in Fig.~\ref{fig:GWnu_scheme}). We can simply perform a temporal-coincidence analysis between these lists. The joint coincidences are defined as ``CCSN candidates''. The statistical significance of such candidates can be estimated by combining the FAR between both lists. To estimate the $\mathrm{FAR_{glob}}$ threshold, we take the same requirements\footnote{Note that this requirement may change in subsequent SNEWS updates.} of SNEWS \cite{Antonioli2004} in the neutrino sub-network, which is the $\mathrm{FAR_\nu} \leq 1/100\,\mathrm{year}$ and a coincidence window $w_{\nu}=10$ seconds, thus the threshold for the GW analysis is $\mathrm{FAR_{GW}}\leq 864$ per day. The coincidence windows in the LEN analysis and in the global network could be in principle very different. In our work, we define a ``detection'' in a network when $\mathrm{FAP}\geq 5\sigma$\footnote{$5\sigma\approx 5.7\times 10^{-7}$}.

\section{Results \label{sec:results}}



Here, we focus only on the multimessenger step. Other results can be seen in \cite{Halim2021}. The first row of Tab.~\ref{tab:dimkamlvd} shows us the comparison of efficiency between the 1-parameter and the 2-parameter method. This gives us an additional {$\sim12\%$} of signals analysed by the 2-parameter method that otherwise are lost by 1-parameter method. 


\begin{table}
    \caption{Efficiency comparison of the 1-parameter ($\eta_\mathrm{1param}$) and the 2-parameter ($\eta_\mathrm{2param}$) method. The first column indicates the specific network of detectors and models (HLV are GW detectors: LIGO-Hanford, LIGO-Livingston, and Virgo; KAM \& LVD are KamLAND and LVD detectors; see \cite{Halim2021}). The second column shows the GW results with its threshold. The third and last columns are the efficiency with $>5\sigma$ significance.}
    \label{tab:dimkamlvd}
    \centering
    {\renewcommand{\arraystretch}{.8}
    \begin{tabular}{|c | c | c | c |} 
    \hline
    Network $\&$ Type  & Recovered   & $\eta_\mathrm{1param}$ & $\eta_\mathrm{2param}$  \\
    of Injections & $\mathrm{FAR_{GW}}<864/\mathrm{d}$ &  $\left[>5\sigma\right]$ & $\left[>5\sigma\right]$  \\
    \hline
    \hline
    HLV-KAM  & 784/2346= &  554/784= &  650/784=  \\
    (Dim2-SN1987A) & 33.4\% & \cellcolor{magenta!30}\textbf{70.7\%} &  \cellcolor{yellow!70}\textbf{82.9\%} \\
    \hline
    HLV-KAM-LVD  & 784/2346= & 776/784= &  784/784=   \\
    (Dim2-SN1987A) & 33.4\% & \cellcolor{magenta!30}\textbf{99.0\%}  &  \cellcolor{yellow!70}\textbf{100\%}  \\
    \hline
    \end{tabular}
    }
    \end{table}


There are 2346 GW injections performed and analyzed with the cWB GW-pipeline, and only 784 of them pass $\mathrm{FAR_{GW}<864/day}$. These candidates' significances are too low to be even considered \textit{sub-threshold} detections. These triggers are then used for coincidences with the list of LEN clusters. Among these multimessenger candidates, {554} pass $5\sigma$ ($\sim 71\%$ of the GW triggers) with the standard 1-parameter method. Whereas, the new 2-parameter method gives us 110 more recovered signals, which increase the efficiency from the GW triggers to $\sim 83\%$ (the first row of Tab.~\ref{tab:dimkamlvd}).

\begin{figure}[!ht]
    \centering
    \includegraphics[width=.6\linewidth]{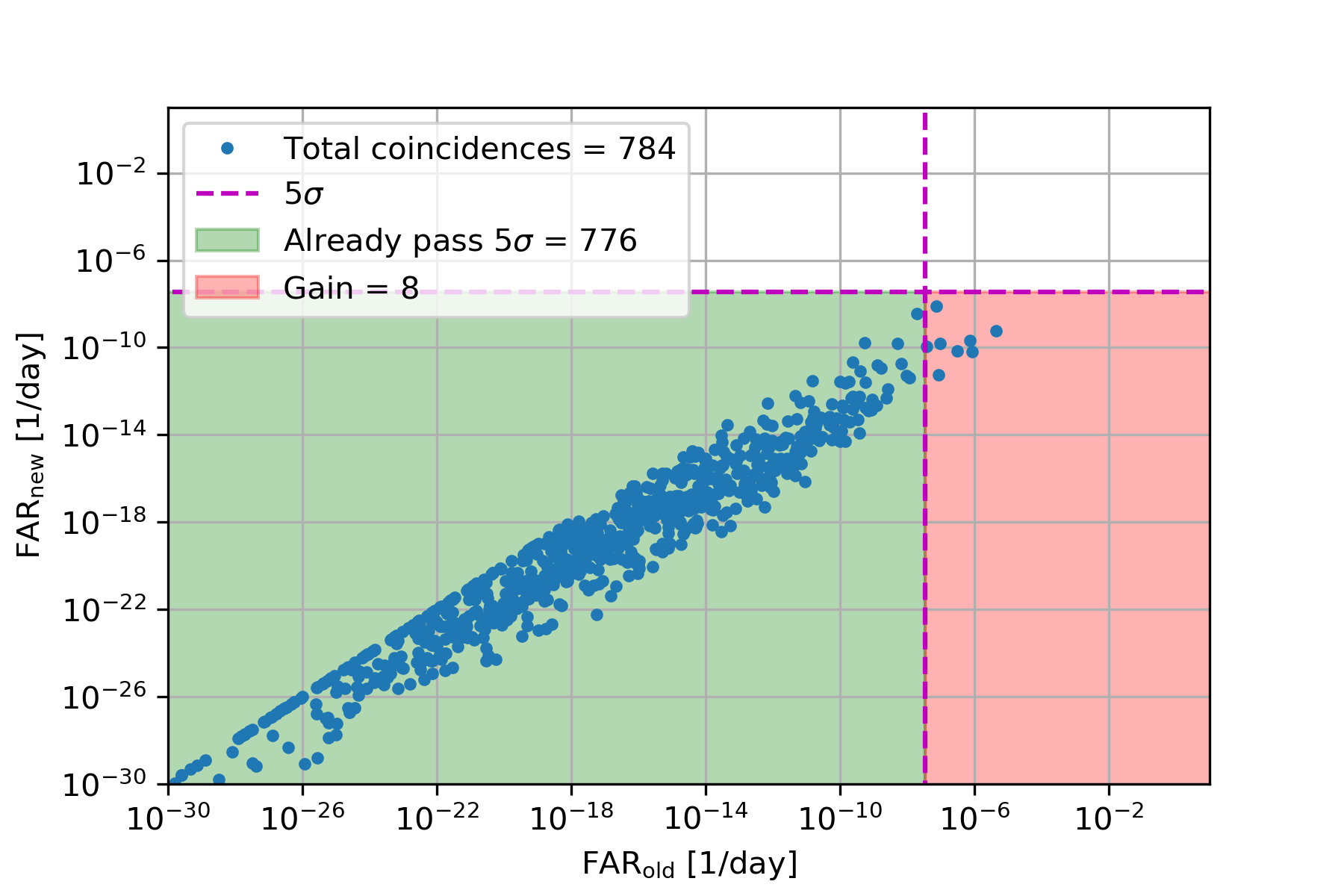}
      \caption{The joint FAR of GW-LEN candidates obtained with the 2-parameter method ($y$-axis) vs the 1-parameter ($x$-axis) with the \textbf{KamLAND-LVD} (SN1987A-model) and HLV (Dim2-model) at 60 kpc.}
      \label{fig:gw_1987_60_tri}
    \end{figure}

Analogously, we extend this method involving also the LVD detector. The results can be seen in Fig.~\ref{fig:gw_1987_60_tri} and summarised in the second row of Tab.~\ref{tab:dimkamlvd}. The improvement of the 2-parameter is less evident. Clearly, the efficiency cannot go beyond its maximum value of $33.4\%$, i.e., all the GW triggers from cWB are coincident LENs and they pass $>5\sigma$.

\section{Conclusion \label{sec:conclusion}}

In this proceeding, we have given a timely introduction of our proposed method for multimessenger search of CCSNe via GWs and {LENs}; more details will be provided in the forthcoming paper \cite{Halim2021}. The multimessenger campaign between GWs and LENs with the proposed method can gain the overall efficiency. Due to the higher sensitivity in LEN detectors, we can also do a targeted search in the (hopefully near) future to search for GWs from CCSNe

\section*{Acknowledgement}
The authors gratefully acknowledge the support of the NSF for the provision of computational resources. The work of GP is partially supported by the research grant number 2017W4HA7S ``NAT-NET: Neutrino and Astroparticle Theory Network'' under the program PRIN 2017 funded by the Italian Ministero dell'Istruzione, dell'Universit\`{a} e della Ricerca (MIUR).


\end{document}